\documentclass[%
superscriptaddress,
amsmath,amssymb,
aps,
twocolumn,
]{revtex4-2}

\usepackage{hyperref}
\hypersetup{colorlinks,allcolors=blue}
\usepackage{graphicx}
\usepackage{bm}
\usepackage[all]{hypcap} 
\usepackage{color}
\usepackage{siunitx}

\begin{document}

\title{Identification of a Critical Doping for Charge Order Phenomena in Bi-2212 Cuprates via RIXS}

\author{Haiyu Lu}
\affiliation{Stanford Institute for Materials and Energy Sciences, SLAC National Accelerator Laboratory, 2575 Sand Hill Road, Menlo Park, California 94025, USA}
\affiliation{Department of Physics, Stanford University, Stanford, California 94305, USA}

\author{Makoto Hashimoto}
\affiliation{Stanford Institute for Materials and Energy Sciences, SLAC National Accelerator Laboratory, 2575 Sand Hill Road, Menlo Park, California 94025, USA}

\author{Su-Di Chen}
\affiliation{Stanford Institute for Materials and Energy Sciences, SLAC National Accelerator Laboratory, 2575 Sand Hill Road, Menlo Park, California 94025, USA}
\affiliation{Department of Applied Physics, Stanford University, Stanford, California 94305, USA}

\author{Shigeyuki Ishida}
\affiliation{National Institute of Advanced Industrial Science and Technology, Tsukuba, Ibaraki 305-8568, Japan}

\author{Dongjoon Song}
\affiliation{National Institute of Advanced Industrial Science and Technology, Tsukuba, Ibaraki 305-8568, Japan}

\author{Hiroshi Eisaki}
\affiliation{National Institute of Advanced Industrial Science and Technology, Tsukuba, Ibaraki 305-8568, Japan}

\author{Abhishek Nag}
\affiliation{Diamond Light Source, Harwell Campus, Didcot OX11 0DE, United Kingdom}

\author{Mirian Garcia-Fernandez}
\affiliation{Diamond Light Source, Harwell Campus, Didcot OX11 0DE, United Kingdom}

\author{Riccardo Arpaia}
\affiliation{Dipartimento di Fisica, Politecnico di Milano, Piazza Leonardo da Vinci 32, I-20133 Milano, Italy}
\affiliation{Department of Microtechnology and Nanoscience, Chalmers University of Technology, SE-41296 Göteborg, Sweden}

\author{Giacomo Ghiringhelli}
\affiliation{Dipartimento di Fisica, Politecnico di Milano, Piazza Leonardo da Vinci 32, I-20133 Milano, Italy}
\affiliation{CNR-SPIN, Dipartimento di Fisica, Politecnico di Milano, Piazza Leonardo da Vinci 32, I-20133 Milano, Italy}

\author{Lucio Braicovich}
\affiliation{Dipartimento di Fisica, Politecnico di Milano, Piazza Leonardo da Vinci 32, I-20133 Milano, Italy}
\affiliation{European Synchrotron Radiation Facility (ESRF), B.P. 220, F-38043 Grenoble Cedex, France}

\author{Jan Zaanen}
\affiliation{Instituut-Lorentz for theoretical Physics, Leiden University, Niels Bohrweg 2, 2333 CA Leiden, The Netherlands}

\author{Brian Moritz}
\affiliation{Stanford Institute for Materials and Energy Sciences, SLAC National Accelerator Laboratory, 2575 Sand Hill Road, Menlo Park, California 94025, USA}

\author{Kurt Kummer}
\affiliation{European Synchrotron Radiation Facility (ESRF), B.P. 220, F-38043 Grenoble Cedex, France}

\author{Nicholas B. Brookes}
\affiliation{European Synchrotron Radiation Facility (ESRF), B.P. 220, F-38043 Grenoble Cedex, France}

\author{Ke-Jin Zhou}
\affiliation{Diamond Light Source, Harwell Campus, Didcot OX11 0DE, United Kingdom}

\author{Zhi-Xun Shen}
\affiliation{Stanford Institute for Materials and Energy Sciences, SLAC National Accelerator Laboratory, 2575 Sand Hill Road, Menlo Park, California 94025, USA}
\affiliation{Department of Physics, Stanford University, Stanford, California 94305, USA}
\affiliation{Geballe Laboratory for Advanced Materials, Stanford University, Stanford, CA, USA}

\author{Thomas P. Devereaux}
\affiliation{Stanford Institute for Materials and Energy Sciences, SLAC National Accelerator Laboratory, 2575 Sand Hill Road, Menlo Park, California 94025, USA}
\affiliation{Geballe Laboratory for Advanced Materials, Stanford University, Stanford, CA, USA}
\affiliation{Department of Materials Science and Engineering, Stanford University, Stanford, California 94305, USA}

\author{Wei-Sheng Lee}
\email{leews@stanford.edu}
\affiliation{Stanford Institute for Materials and Energy Sciences, SLAC National Accelerator Laboratory, 2575 Sand Hill Road, Menlo Park, California 94025, USA}


\begin{abstract}
Identifying quantum critical points (QCPs) and their associated fluctuations may hold the key to unraveling the unusual electronic phenomena observed in cuprate superconductors. Recently, signatures of quantum fluctuations associated with charge order (CO) have been inferred from the anomalous enhancement of CO excitations that accompany the reduction of the CO order parameter in the superconducting state. To gain more insight about the interplay between CO and superconductivity, here we investigate the doping dependence of this phenomenon throughout the Bi-2212 cuprate phase diagram using resonant inelastic x-ray scattering (RIXS) at the Cu $L_3$-edge. As doping increases, the CO wavevector decreases, saturating at a commensurate value of 0.25 {\it r.l.u.}~beyond a characteristic doping $p_{c}$, where the correlation length becomes shorter than the apparent periodicity (4$a_0$). Such behavior is indicative of the fluctuating nature of the CO; and the proliferation of CO excitations in the superconducting state also appears strongest at $p_{c}$, consistent with expected behavior at a CO QCP. Intriguingly, $p_c$ appears to be near optimal doping, where the superconducting transition temperature $T_c$ is maximal. 
\end{abstract}

\maketitle

\section*{Introduction}
Electronic complexity in high-$T_c$ cuprate superconductors manifests in multiple intertwined, competing or cooperating phases, coexisting with or proximal to superconductivity \cite{Fradkin2015, Keimer2015}. Whether these phases undergo quantum phase transitions remains an important question, because if they do, the associated quantum fluctuations may account for the unusual properties of the "normal" state and possibly affect superconductivity \cite{Fradkin2015,Keimer2015,Sachdev2010,Zaanen2004,Kivelson2003, Castellani1995, Castellani1996}. While evidence exists for vanishing order parameters from Kerr rotation \cite{Jing2008}, optical second harmonic spectroscopy \cite{Zhao2017} and x-ray scattering \cite{Blanco2014}, signatures of quantum critical behavior in the cuprates are inferred primarily from transport measurements \cite{TALLON1994,Ramshaw2015,Taillefer2010, Proust2019}, which do not probe directly the order parameter of a quantum phase. Without direct access to an order parameter or excitation spectrum, drawing associations between quantum phases and putative quantum critical fluctuations becomes a difficult task.

Due to its ubiquitous presence in superconducting cuprates, and the intertwined behavior of its order parameter with that of superconductivity, it is of great interest to investigate whether charge order (CO)\cite{Tranquada1995,Howald2003,Ghiringhelli2012YBCO,Chang2012,Comin2014,Hashimoto2014,daSilvaNeto2014,Tabis2014,Uchida2021,Arpaia2019,Arpaia2021,Tranquada2021,Lee2021_review, Wahlberg2021} possesses a quantum phase transition, and in particular one tuned by chemical substitution or doping. The order parameter has been probed directly by neutron and x-ray scattering, as well as STM, but those investigations have provided little direct evidence about the CO excitation spectrum or the presence of quantum critical fluctuations. In principle, quantum fluctuations drastically modify collective behavior; however, these excitations form a ``continuum'' \cite{Sachdev_Book,Kivelson2003}, rather than sharply defined collective modes, often leaving only subtle fingerprints on the spectra. 

In this regard, resonant inelastic x-ray scattering (RIXS) can provide a unique perspective on CO and its excitations. RIXS directly probes the CO order parameter, allowing one to track the momentum distribution of the quasi-elastic peak intensity in the spectrum. It also can highlight the presence of CO excitations, and their interplay with lattice degrees of freedom, through the variation of dispersion and spectral intensity attributed to collective vibrational modes -- phonons. Proportional to the electron-phonon coupling \cite{Devereaux2016}, the RIXS cross-section for phonon excitations also exhibits a Fano-like interference effect between the sharply defined phonon modes and the underlying ``continuum'' charge excitations. Demonstrated previously, the cross-section for the CO excitations themselves can be small, but their Fano fingerprint manifests as an enhancement of the phonon intensity and a softening of the "phonon" dispersion in the RIXS map\cite{Chaix2017, Devereaux2016,Li2020,Huang2021}. The RIXS phonon cross-section is distinct from the phonon self-energy measured using inelastic neutron scattering (INS) or non-resonant inelastic x-ray scattering (IXS); thus, the dispersion and spectral lineshape (width and intensity) of the RIXS phonons will deviate from that of the phonons measured using INS and IXS, in particular due to the influence from the underlying charge excitations.

Interestly, temperature dependent studies of RIXS phonon excitations in nearly optimally doped Bi$_2$Sr$_2$CaCu$_2$O$_{8+\delta}$ \cite{Lee2021} and La$_{2-x}$Sr$_x$CuO$_4$ \cite{Huang2021} have captured unexpected behaviors, directly attributable to the influence of CO quantum fluctuations. Cooling through the superconducting transition, the CO order parameter decreases as the superconducting gap grows. However, in contrast to the behavior ascribed to a Landau phase transition, where the spectral weight associated with CO excitations should be proportional to the amplitude of the order parameter, the Fano effects and influence from CO excitations become more pronounced in the RIXS phonon cross-section. This seemingly anomalous temperature dependence below $T_c$ portends the approach to a QCP \cite{Lee2021,Huang2021}:  the opening of a superconducting gap reduces the dissipation for quantum fluctuations; quantum fluctuations are restored at low temperatures, which effectively moves the system towards the QCP \cite{Lee2021}. Such a scenario accounts for the apparent contradiction, in which the CO order parameter decreases while the Fano effects and underlying spectral weight associated with the continuum increase, here, restored by the growth of quantum fluctuations upon entering the superconducting state. However, results have been reported only at a singular point, near optimal doping. To gain further insight and support for such a scenario requires tracking the anomalous behavior as a function of doping throughout the cuprate phase diagram.   

In this article, we report on the doping dependence of the anomalous enhancement in the influence of CO excitations below $T_c$. High resolution RIXS measurements both at $T_c$ and 15 K, the lowest achievable temperature of our instrument, reveal clear signatures of CO quantum fluctuations. We report data taken on samples with eight different doping concentrations, covering the antiferromagnetic, under-, optimally-, and over-doped (AFM, UD, OP, and OD) regions across the cuprate phase diagram. We find that the anomalous enhancement of CO excitations below $T_c$ exhibits a dome shape as a function of doping with a maximum at $p_c$ located between $p$ = 0.13 and 0.16 holes/Cu. For $p>p_c$, the CO order parameter extracted from the quasi-elastic RIXS spectrum also exhibits qualitative changes: i) the CO wavevector saturates at 0.25 {\it r.l.u.}~and ii) the CO correlation length becomes shorter than its periodicity. Overall, these experimental findings are consistent with the existence of a CO QCP in the vicinity of $p_c$ in the Bi-2212 doping phase diagram.

\section*{Experiment details}
High-quality Bi$_{1.7}$Pb$_{0.4}$Sr$_{1.7}$CaCu$_2$O$_{8+\delta}$ (Bi-2212) single crystals were grown by floating-zone methods. A broad range of oxygen dopings were achieved by a subsequent annealing process. In this study, we systematically studied OP96 ($T_c$=96K), OD90 ($T_c$=90K), OD83 ($T_c$=83K), and OD70 ($T_c$=70K) samples. We have also studied underdoped UD70 samples ($T_c$=70K) which were annealed from high quality Bi$_{2.1}$Sr$_{1.9}$CaCu$_2$O$_{8+d}$ single crystals. The superconducting transition was determined by the onset temperature $T_c$ measured with the AC magnetic susceptibility module of a Physical Property Measurement System from Quantum Design. The hole doping concentration was estimated using the empirical parabolic function \cite{Presland1991}, $T_c=T_{c, max}[1-82.6(p-0.16)^2]$, where $T_{c, max}$ = 96 K for the batch of crystals studied in this experiment. By using this formula, we are assuming that the optimal doping concentration is $p_{op} = 0.16$ holes/Cu. For completeness, we have also studied a non-superconducting and insulating antiferromagnetic (AFM) compound, Bi$_{1.6}$Pb$_{0.4}$Sr$_2$YCu$_2$O$_{8+\delta}$, whose doping concentration cannot be determined using the empirical function, but is expected to be nearly undoped. All crystals were selected with flat surfaces and a sharp Laue pattern.

The Cu $L_3$-edge RIXS experiments on OP96, OD90, OD83, and OD70 samples were conducted in a single beamtime using the RIXS spectrometer at the I21 RIXS beamline of the Diamond Light Source (DLS) in the United Kingdom \cite{Zhou2022}. The data of AFM and UD70 were obtained using the ERIXS spectrometer at the ID32 beamline of the European Synchrotron Radiation Facility (ESRF) in France \cite{BROOKES2018}, where preliminary measurement on OD70 was also conducted. High-resolution RIXS data were measured with a combined energy resolution of 37 meV at DLS and 44 meV at ESRF, respectively. For all superconducting compounds, momentum dependent RIXS maps were measured both at $T_c$ and well below (15 K), whereas the non-superconducting AFM sample was measured only at 15 K. See the Supplemental Material for details about the RIXS measurements. For all data shown here, the electronic structure of Bi-2212 has been treated as quasi-two-dimensional and momentum dependent data have been denoted as a function of the projected in-plane momentum transfer $q_{\parallel}$ along the Cu-O bond direction. More details about the scattering geometry can be found in the Supplementary Material.

\section*{Results}
Figure ~\ref{Figure 1} shows the RIXS intensity map for the UD70 sample, simultaneously recording the CO peak in the quasi-elastic region and the RIXS intensity for phonon excitations. The quasi-elastic map in Fig.~\ref{Figure 1}(b) has been obtained by subtracting a fit of the RIXS phonon intensity from the full RIXS spectrum (see the Supplementary Material for details), with the momentum distribution curve (MDC) obtaind by integrating the map within an energy window of $\pm25$ meV. The CO peak can be seen clearly, centered at $Q_{CO}=0.275$ {\it r.l.u.} (indicated by the black dotted line). To better visualize the RIXS phonon spectra, the fitted quasi-elastic portion of the RIXS maps was subtracted from the raw data with the result plotted in Fig.~\ref{Figure 1}(c). Here, at higher energies around 60 meV, a branch of excitations, the RIXS phonons, can be identified, whose peak position are indicated by white markers. Consistent with recent high-resolution RIXS experiments at the Cu $L_3$-edge in the literature \cite{Chaix2017, Lee2021}, the RIXS phonons not only exhibit an apparent softening at the CO wavevector ($Q_{CO}$) but also show a ``funnel''-shaped spectral weight emanating from $Q_{CO}$ with a non-monotonic momentum distribution of the integrated intensity (see the lower panel of Fig.~\ref{Figure 1}(c)), whose maximum (black arrow) occurs at $Q_{A}>Q_{CO}$. These anomalies in the RIXS map were attributed to Fano-like interference between the phonons and underlying dispersive CO excitations \cite{Chaix2017, Lee2021}, which form a continuum.

As reference, we turn to the AFM sample, which is nearly undoped with antiferromagnetic (AFM) order. As expected, neither the raw data (Fig.~\ref{Figure 1}(d)) nor the quasi-elastic map (Fig.~\ref{Figure 1}(e)) hint at the presence of CO, indicating that CO in Bi2212 emerges at a finite doping concentration in the heavily underdoped regime of the phase diagram, as in YBCO \cite{Blanco2014, Jang2018}. As one sees in Fig. ~\ref{Figure 1}(f), the RIXS phonon intensity exhibits neither an apparent dispersion softening nor a non-monotonic distribution of integrated intensity as a function of momentum, due to the absence of CO. In fact, the spectral weight of phonon excitations increases monotonically with increasing $q_{\parallel}$, as expected from the momentum dependence of the electron-bond-stretching-phonon coupling \cite{Devereaux2016}. The stark contrast in behavior of phonon excitation spectral weight between the AFM sample where CO is absent and UD70 sample (and Refs.~\cite{Chaix2017} and \cite{Lee2021}) where CO is present unambiguously confirms the sensitivity of RIXS phonons to the CO excitations, making RIXS an effective tool for detecting a putative quantum phase transition associated with CO.

\begin{figure}
	\centering
	\includegraphics[width=1.0\columnwidth]{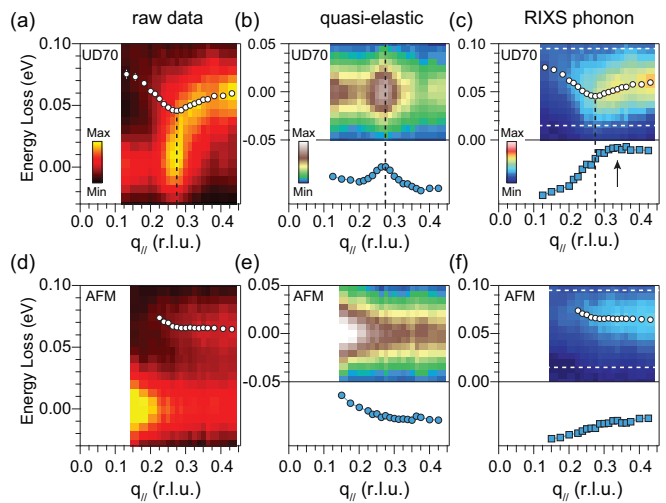}
	\caption[UD70 vs AFM at 15K]{\label{Figure 1} RIXS intensity maps of (a) UD70 samples as a function of energy loss and projected in-plane momentum along the Cu-O bond direction measured at 15 K. White markers are fitted peak positions of the RIXS phonon intensity. The black dashed line indicates the CO wavevector $Q_{CO}$ in the UD70 sample. (b) RIXS quasi-elastic map obtained by subtracting the fitted RIXS phonon intensity from the spectrum. The MDC (lower panel) was obtained by integrating the quasi-elastic map between ($\pm25$ meV). (c) RIXS phonon map obtained by subtracting the fitted quasi-elastic peak from the raw data. The MDC (lower panel) shows the integrated intensity between the white dashed lines in the top panel (from 15 to 95 meV). The black dashed line and arrow indicate $Q_{CO}$ and the maximum of the curve, respectively. (d), (e), and (f) show the raw RIXS intensity map, quasi-elastic map, and the RIXS phonon map for the nearly-undoped AFM sample.}
\end{figure}

\begin{figure*}
	\centering
	\includegraphics[width=2\columnwidth]{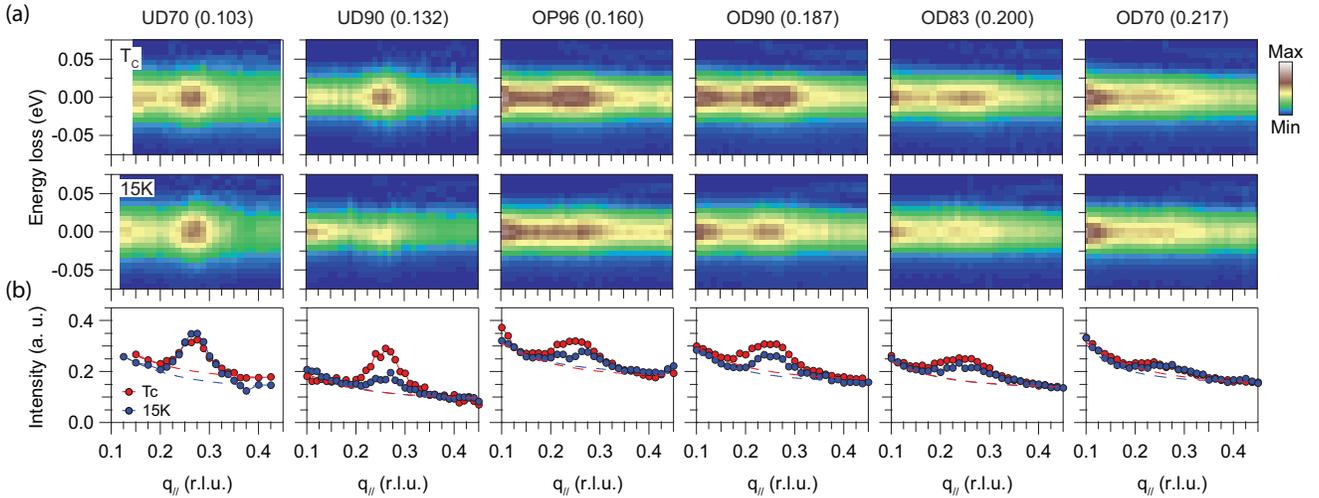}
	\caption[Quasi-elastic maps from Bi-2212 taken at 15 K and $T_c$]{\label{Figure 2} (a) Quasi-elastic maps of Bi-2212 taken at $T_c$ and 15 K, showing the temperature and doping dependence of the CO peak. All the data have been normalized to the $dd$ excitations (the integrated intensity in the RIXS maps between 1 and 4 eV energy loss). The hole-doping of each sample is indicated by the number in the parenthesis at the top of each panel. (b) The CO peak profile obtained by integrating quasi-elastic RIXS maps within the energy loss window of $\pm$ 25 meV. The dashed curves are background used in fitting the CO peak. The data from UD90K has been adapted from Ref.~\cite{Lee2021}. 
	}
\end{figure*}

To study the temperature and doping dependence of CO and its excitations, we have compiled momentum dependent RIXS maps for all the other dopings at both 15 K and $T_c$. The intensity of the RIXS maps has been normalized to the $dd$ excitation peak area for different samples (see the Supplementary Material). The UD90 data has been adapted from Ref.~\cite{Lee2021} for comparison. We focus first on the behavior of CO in the quasi-elastic region, which can be inferred from the quasi-elastic map and the MDCs as shown in Fig.~\ref{Figure 2}. At first glance, the CO exhibits more-or-less the expected doping dependence. In the under-doped regime, the CO peak is well defined with a peak position corresponding to the CO wavevector $Q_{CO}$. With increasing doping, the CO peak shifts and broadens. In the OD70 sample, the most overdoped sample in our study, while a signature of CO still may be resolvable, the broad peak profile indicates that CO is no longer a well defined state. Both $Q_{CO}$ and the peak width ($\sigma_{CO}$) can be determined by fitting the peak profile to a Gaussian function plus a background arising from specular refection at zero momentum, which we mimic by a Lorentzian function in the fit. 

\begin{figure}[h]
	\centering
	\includegraphics[width=\columnwidth]{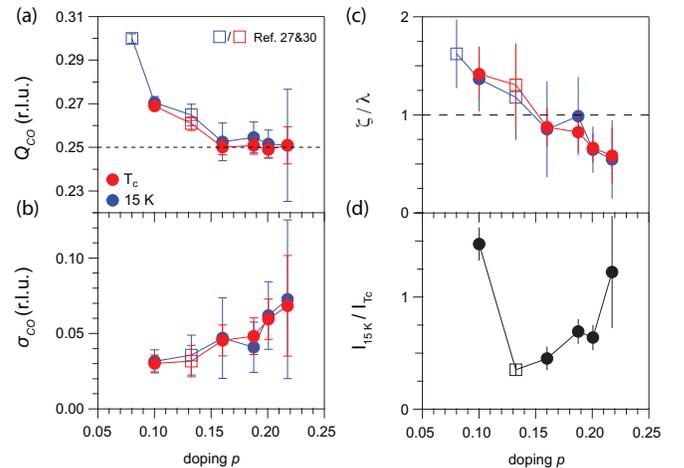}
	\caption[CO profiles in Bi-2212 cuprates]{\label{Figure 3} The doping dependence of (a) the CO ordering wavevector, $Q_{CO}$ and (b) the CO peak width $\sigma_{CO}$, represented by the standard deviation of the Gaussian peak used to fit the CO peak profile. The UD45(0.08) and UD90(0.132) data are reproduced from Refs.~\cite{Chaix2017} and \cite{Lee2021}, plotted as open square markers for comparison. (c) A summary of the ratio between the CO periodicity $\lambda = a / Q_{CO}$ and the correlation length $\zeta = a / (2\pi \sigma_{CO})$, where $\sigma_{CO}$ is shown in (b) and $a$ denotes the lattice constand $a = 3.89$ \AA. (d) The CO peak intensity ratio $I_{15 K} / I_{T_c}$ between 15 K and $T_c$. The CO peak intensity is defined as the CO peak profile height, subtracting the fitted background intensity at $Q_{CO}$.}
\end{figure}

The results from the fit are summarized in Fig.~\ref{Figure 3}. While the $\sigma_{CO}$ monotonically increases with doping, $Q_{CO}$ exhibits an intriguing doping evolution. In the under-doped regime, $Q_{CO}$ is incommensurate and monotonically decreases with increasing doping. Near the optimal doping $p_{op}=0.16$ holes/Cu, $Q_{CO}$ begins to saturate at a commensurate value of 0.25 {\it r.l.u.} in the overdoped regime. This indicates that CO may undergo a qualitative change at a doping concentration $p_c$, which is located between 0.13 holes/Cu and optimal doping $p_{op}$. Interestingly, by comparing the CO periodicity ($\lambda$) and the CO correlation length ($\zeta$) as shown in Fig. \ref{Figure 3}(c), $\zeta / \lambda$ becomes less than 1 for doping concentrations $p>p_c$, indicating that the correlation length is shorter than the apparent CO periodicity. In other words, the CO is not well defined, and the charge order fluctuating become dominant in this region of phase diagram.

Notably, the temperature dependence of CO below $T_c$ reveals a subtle but intriguing doping dependent behavior. While CO is suppressed in the superconducting state in samples near optimal doping (\textit{e.g.} UD90, OP96, and OD90K), CO appears to be insensitive to superconductivity in the under-doped (\textit{e.g.} UD70) and over-doped (\textit{e.g.} OD70) regions. To better quantify the CO suppression, Fig.~\ref{Figure 3}(d) shows the CO order peak intensity ratio between $T_c$ and 15 K, \textit{i.e.} $I_{15 K}/I_{Tc}$. In other words, the CO suppression due to superconductivity is not monotonic as a function of doping, indicating that the origin of the suppression is likely more complex than the apparent phase competition between CO and superconductivity. Interestingly, this suppression is most pronounced near $p_c$, which more or less coincides with the doping concentration where the $Q_{CO}$ saturate to 0.25 {\it r.l.u.}~(Fig. \ref{Figure 3}).

\begin{figure*}
	\centering
	\includegraphics[width=2\columnwidth]{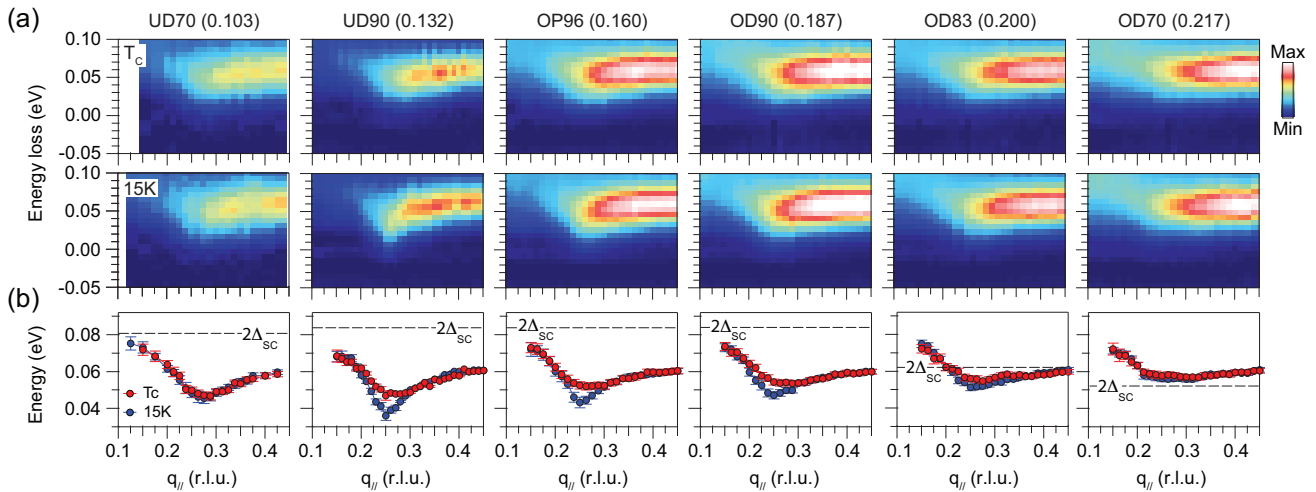}
	\caption[Raw data of the RIXS phonon intensity in Bi-2212 cuprates at 15 K and $T_c$]{\label{Figure 4} (a) RIXS phonon maps from Bi-2212 taken at $T_c$ and 15K. The intensity is normalized to the $dd$ excitations (integrated peak area from 1 to 4 eV energy loss). The hole-doping of each sample is indicated by the number in the parenthesis at the top of each panel. (b) Doping dependence of the RIXS phonon dispersion at $T_c$ and 15K. The error bar represents the standard deviation from the fitting. Horizontal dashed lines indicate the superconducting gap ($2\Delta_{SC}$) estimated from Ref. \cite{Vishik2012} and \cite{He2018} for the associated doping.}
\end{figure*}

Next, we examine the RIXS phonon anomalies near $Q_{CO}$, which are related to the underlying CO excitations. Figure~\ref{Figure 4}(a) shows the RIXS phonon maps at $T_c$ and 15 K, well below $T_c$. At $T_c$, the dispersion softening near $Q_{CO}$ is most apparent in the underdoped regime and becomes less and less pronounced with increased doping, essentially following the diminishing and broadening CO order parameter in the quasi-elastic region (Fig. \ref{Figure 2}). Yet, the evolution of the dispersion in the superconducting state at 15 K exhibits an intriguing doping dependence. As shown in Fig.~\ref{Figure 4}(b), in the UD70 and OD70 samples, whose CO peak profiles show little temperature dependence at and below $T_c$, the phonon dispersion also remains roughly the same after entering the superconducting state. In other words, the CO excitations to which the RIXS phonons couple do not exhibit noticeable changes when the system becomes superconducting in the significantly under- and over-doped regimes. In stark contrast, the samples near optimal doping (UD90, OP96, and OD90) show strong phonon softening near $Q_{CO}$, which becomes even more pronounced in the superconducting state. The ``continuum" of CO excitations must increase in the superconducting state, despite of the fact that CO in the quasi-elastic region decreases (see Fig.~\ref{Figure 2}). The enhanced phonon softening near $Q_{CO}$ reflects the anomalous enhancement of CO excitations in the superconducting state. As summarized in Fig. \ref{Figure 5}, it exhibits a dome-shaped doping dependence, in connection with the superconductivity-induced CO suppression in the quasi-elastic peak (Fig. \ref{Figure 3}(d)). We note that the non-monotonic doping dependence of the phonon softening at $Q_{CO}$ is evident already in the raw data, as shown in the Supplementary Material Fig. 3.

As argued previously \cite{Lee2021}, the enhancement of CO excitations in the superconducting state is due to the weakening of electronic damping by the opening of the superconducting gap ($2\Delta_{SC}$). This provides a similar damping reduction in UD70, UD90, OP96, and OD90 samples, since $2\Delta_{SC}$ is well-above the RIXS phonon energy near $Q_{CO}$ and rather doping insensitive in the under- and optimally-doped regimes (Fig. \ref{Figure 4}(b) and Ref.\cite{Vishik2012,He2018}). Thus, the dome-shaped doping dependence of the RIXS phonon shift (Fig. \ref{Figure 5}) in this portion of the phase diagram should reflect the intrinsic doping dependence of the CO excitation spectral weight. Though, at higher dopings (OD83 and OD70) where 2$\Delta_{SC}$ becomes comparable to the RIXS phonon energy, the weakening of the RIXS phonon enhancement in the superconducting state results from both the reduction of the CO excitation spectral weight but also the smaller 2$\Delta_{SC}$.

\section*{Discussion and conclusions}

\begin{figure}
	\centering
	\includegraphics[width=\columnwidth]{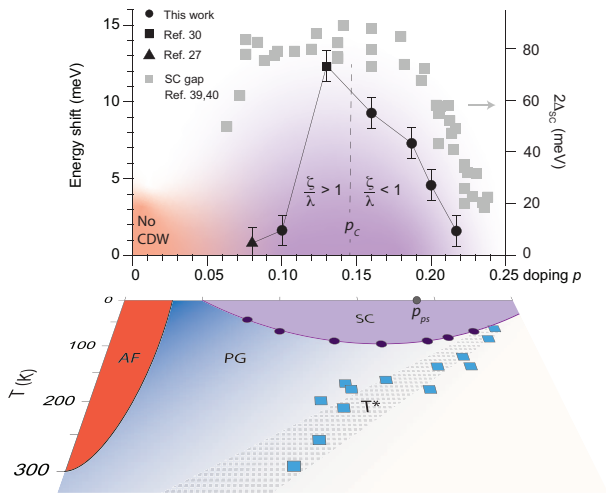} 
	\caption[QCP]{\label{Figure 5} A summary of the RIXS phonon energy shift at $Q_{CO}$ between $T_C$ and 15 K as a function of doping, except for the data point of $p$ = 0.08, which is deduced from the energy shift between 300 K and 15 K from Ref. \cite{Chaix2017}, presenting a upper bound for superconductivity-induced energy shift. The $p_c$ is the characteristic doping where the ratio of $\zeta/\lambda = 1$. The values of superconducting gap (SC gap, 2$\Delta_{SC}$) were adapted from Ref. \cite{Vishik2012, He2018}. The Bi2212 phase diagram adopted from Ref. \cite{Vishik2012} is also shown in the lower part of the figure. The grey marker at $p_{ps} = 0.19$ denotes the position of the putative critical point of pseudogap state.}
\end{figure}

Whether CO in cuprates arises from Fermi surface nesting or strong electronic correlations has been a topic of considerable debate \cite{Comin2016}. Our observation of $Q_{CO}$ saturating at a commensurate value of 0.25 \textit{r.l.u.} (\textit{i.e.} 4$a_0$ periodicity) in the over-doped regime indicates that the formation of CO is unlikely due to any Fermi surface nesting because the nesting vector would continuously decrease in such scenarios. We note that a similar conclusion has also been reached for the CO in La-based cuprates \cite{Miao2019}. That $Q_{CO}$ locks in to a commensurate value should have been a stabilizing force for CO; however, we observe a diminishing CO that becomes broader and weaker in the overdoped regime. This observation could be reconciled through discommensuration, an effect which can yield an incommensurate $Q_{CO}$ in reciprocal space (\textit{i.e.} x-ray scattering probes) from locally commensurate 4$a_0$ CO domains in real space, as seen by STM \cite{Mesaros2016}. Since this effect is highly sensitive to the presence of disorder, it should decrease with increasing hole doping; thus, $Q_{CO}$ seen by x-ray scattering should approach the commensurate value of 0.25 \textit{r.l.u.}, as shown in our results. However appealing such a scenario may be, it is incomplete, since the CO excitations and the associated RIXS phonon anomalies should not be affected by discommensuration, but rather they should occur at 0.25 \textit{r.l.u.} through out the entire phase diagram, which seems inconsistent with the data in the underdoepd regime \cite{Chaix2017}. This reveals a subtle but intriguing apparent paradox that we present as a challenge to the community.

The essence of the CO behavior throughout the Bi-2212 phase diagram is summarized in Fig.~\ref{Figure 5}. CO is absent in our AFM sample, suggesting that it emerges at a finite doping concentrations in the heavily-underdoped region. With increased doping, the CO wave-vector shifts, the peak width increases, and the CO eventually becomes fluctuating for doping concentrations beyond some critical value $p_c$. Beyond this point, the apparent CO periodicity $\lambda$ settles at a commensurate value of 4$a_o$ and exceeds the correlation length $\zeta$ ({\it i.e.}~$\zeta / \lambda < 1$). Remarkably, comparing the temperature dependence between $T_c$ and 15 K ({\it i.e.}~deep in the superconducting state) also suggests the existence of such a characteristic doping concentration. Near $p_c$, the most pronounced suppression of the CO order parameter occurs upon entering the superconducting state. Concomitantly, we observe the strongest enhancement of the CO excitations in the superconducting state, which manifest as the RIXS phonon energy difference at $Q_{CO}$ between $T_c$ and 15 K (Fig. \ref{Figure 5}). Therefore, taken together, these observations are consistent with a scenario which identifies $p_c$ with the closest approach to a CO QCP, where the CO should vanish and become fluctuating and the spectral weight assocaited with the CO continuum would become the most pronounced. We note that signatures of CO for $p>p_c$ remain detectable in the quasi-elastic spectrum of our data, likely indicating the important role that disorder plays in preventing a sharp quantum phase transition.

Previously, a number of experiments, including ARPES \cite{Vishik2012, Chen2019}, STM \cite{Fujita2014}, and transport measurements \cite{Tallon2003a,Tallon2003b}, have reported various anomalies across a characteristic doping concentration $p_{ps}\sim0.19$, which has been associated with the termination of the pseudogap phenomena. In particular, a recent higher resolution ARPES study shows that the pseudogap abruptly disappears at $p_{ps}\sim0.19$, like a first order phase transition as a function of doping concentration \cite{Chen2019}. The fact that the we do not observe any abrupt changes in the CO behavior across $p_{ps}$, indicates that CO and its excitations are not directly associated with the pseudogap phenomena. We note that the persistence of CO beyond $p_{ps}$ in the overdoped region of the phase diagram also has been reported in La-based cuprates \cite{Miao2021} and single layer Bi-2201 cuprates \cite{Peng2018}, lending further support to the disconnect between CO and the $p_{ps}$ anomalies.

Finally, our results suggest that the CO QCP appears to coincide with optimal doping, raising an intriguing question about the role of CO quantum fluctuation in the mechanism for superconductivity. We note that as shown in the Fig. \ref{Figure 5}, 2$\Delta_{SC}$ is essentially doping independent in the range of $0.08<p<0.18$, despite the strongest CO excitation spectral weight at $p_c$. Thus, we conclude that the CO excitations do not play an obvious role in Cooper pairing. Whether the CO excitations near a QCP could boost $T_C$ or whether the $p_c$ being near optimal doping is merely coincidental warrant future investigation.

\section*{Acknowledgement}
This work is supported by the U.S. Department of Energy (DOE), Office of Science, Basic Energy Sciences, Materials Sciences and Engineering Division, under contract DE-AC02-76SF00515. We also acknowledge the Diamond Light Source, UK for providing beamtime at the I21-RIXS beamline under the Proposal MM22009. We also acknowledge the European Synchrotron Radiation Facility, France for providing beamtime at the ID32-RIXS beamline. The crystal synthesis work was supported by the JSPS KAKENHI (No. JP19H05823).

%

\end{document}